\documentclass[reprint, aps, pra]{revtex4-2}
\usepackage{amsmath}
\usepackage{amsfonts}
\usepackage{physics}
\usepackage{graphicx}
\usepackage[colorlinks=true, linkcolor=blue, citecolor=green, urlcolor=blue]{hyperref}
\usepackage{xcolor}
\usepackage[english]{babel}
\usepackage{tikz}

\newcommand{\ee}{\mathrm{e}}
\newcommand{\im}{\mathrm{i}}

\newcommand{\lblad}[1]{\mathcal{L}[#1]}
\newcommand{\wc}{\omega_\mathrm{c}}

\newcommand{\down}{\downarrow}
\newcommand{\up}{\uparrow}
\newcommand{\eV}{\,\mathrm{eV}}
\newcommand{\fs}{\,\mathrm{fs}}
\newcommand{\opr}[1]{\hat{#1}}  

\begin{document}
\title{Theory of dynamical superradiance in organic materials}

\author{Lukas~Freter$^1$}
\author{Piper~Fowler-Wright$^2$}
\author{Javier~Cuerda$^1$}
\author{Brendon~W.~Lovett$^2$}
\author{Jonathan~Keeling$^2$}
\email{jmjk@st-andrews.ac.uk}
\author{Päivi~Törmä$^{1}$}
\email{paivi.torma@aalto.fi}
\affiliation{{$^1$}Department of Applied Physics, Aalto University School of Science, FI-00076 Aalto, Finland\\
{$^2$}SUPA, School of Physics and Astronomy, University of St Andrews, St Andrews KY16 9SS, United Kingdom}

\date{\today}
\begin{abstract}
We develop the theory of dynamical superradiance---the collective exchange of energy between an ensemble of initially excited emitters and a single-mode cavity---for organic materials where electronic states are coupled to vibrational modes. We consider two models to capture the vibrational effects: first, vibrations treated as a Markovian bath for two-level emitters, via a pure dephasing term in the Lindblad master equation for the system; second, vibrational modes directly included in the system via the Holstein--Tavis--Cummings Hamiltonian. By exploiting the permutation symmetry of the emitters and weak U(1) symmetry, we develop a numerical method capable of exactly solving the Tavis-Cummings model with local dissipation for up to 140 emitters. Using the exact method, we validate mean-field and second-order cumulant approximations and use them to describe macroscopic numbers of emitters. We analyse the dynamics of the average cavity photon number, electronic coherence, and Bloch vector length, and show that the effect of vibrational mode coupling goes beyond simple dephasing. Our results show that superradiance is possible in the presence of vibrational mode coupling; for negative cavity detunings, the vibrational coupling may even enhance superradiance. We identify asymmetry of the photon number rise time as a function of the detuning of the cavity frequency as an experimentally accessible signature of such vibrationally assisted superradiance.
\end{abstract}

\maketitle


\section{Introduction}
\begin{figure}
    \centering
    \begin{tikzpicture}
    \node[anchor=south west] (a) at (0,0) {\includegraphics[width=0.65\linewidth]{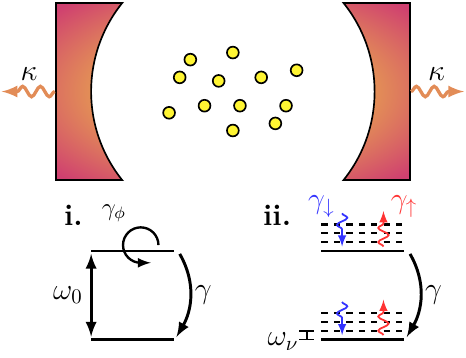}};
    \node[anchor=south west] (b) at (6,0.5) {\includegraphics[width=0.25\linewidth]{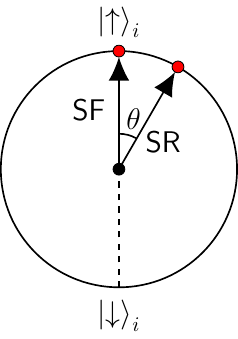}};
    \node[font=\bfseries, shift={(0.35,-.1)}] at (a.north west) {(a)};
    \node[font=\bfseries, shift={(6,-.1)}] at (a.north west) {(b)};
    \end{tikzpicture}
    \caption{(a) Schematic of the dynamical SR system. Emitters (yellow circles) are placed inside a single-mode cavity, which is subject to photon losses at a rate $\kappa$. The internal structure of the emitters follows either (i) the TC model or (ii) the HTC model. For the TC model, the emitters are ideal two-level systems with energy splitting $\omega_0$, and 
    are subject to non-radiative decay at rate $\gamma$ and pure dephasing at rate $\gamma_\phi$. For the HTC model, the emitters are dressed by vibrational levels with spacing $\omega_\nu$, with additional vibrational thermalization  (rates $\gamma_\up$ and $\gamma_\down$).
    (b) Illustration of SF and SR initial conditions 
    according to the polar angle \(\theta\) of the state of each two-level system.
    }
    \label{fig:system-initial-conditions}
\end{figure}

The term superradiance (SR), first introduced by Dicke in 1954~\cite{dicke_coherence_1954}, refers to the collective spontaneous emission of $N$ closely spaced quantum emitters in free space.
Due to the interaction of the emitters with a common light field, a macroscopic dipole moment builds up during the emission process, resulting in a peak emission rate which scales as $N^2$, compared to $\sim N$ for independent emitters. 
For an introduction to SR, see Refs.~\cite{gross_superradiance_1982, cong_dicke_2016};
we also highlight recent interest in the phenomenon of SR in extended systems of different geometries~\cite{carmichael_quantum_2000, clemens_collective_2003, clemens_shot--shot_2004,scully_directed_2006, zoubi_metastability_2010,bhatti_superbunching_2015,shahmoon_cooperative_2017,masson_many-body_2020, masson_universality_2022,rubies-bigorda_superradiance_2022, rubies-bigorda_characterizing_2023, mok_dicke_2023, masson_dicke_2024, holzinger_collective_2025,holzinger_beyond_2025}. In the special case where the emission process starts from an incoherent, fully inverted ensemble of emitters, the process has also been called superfluorescence (SF). 
We adopt the convention~\cite{cong_dicke_2016} that SR refers to any case where the initial state has some non-zero coherence, and SF to the specific case where the initial state has no coherence [Fig~\ref{fig:system-initial-conditions}(b)].

While the original SR and SF, as discussed above, occur for emitters in free space, there are several closely related phenomena that occur for emitters in a cavity.
The first, ``dynamical SR'', describes the dynamics of initially excited emitters coupling collectively to a single cavity mode~\cite{abate_exakte_1964,bonifacio_coherent_1970,barnett_collective_1984,seke_influence_1989,seke_collapse_1989,eastham_new_2007}. Dynamical SR is the subject of this paper.
The second is the superradiant phase transition of the Dicke model~\cite{hepp_superradiant_1973, wang_phase_1973, baumann_dicke_2010, kirton_introduction_2019, roses_dicke_2020}, which describes a steady-state phenomenon of the many-emitter--cavity system.
The third is the idea of the ``superradiant laser''~\cite{haake_superradiant_1993, meiser_prospects_2009,bohnet_steady-state_2012}, which can be considered either as corresponding to the original Dicke SR phenomena but continuously refreshed to provide a steady state, or considered as using a bad cavity to synchronise emitters in an analog of Huygen's clocks.

If the emitters coupled to the cavity mode are treated as ideal two-level systems, and one makes the rotating wave approximation, then the combined light-matter system can be described with the Tavis--Cummings (TC) model~\cite{tavis_exact_1968}. 
In this case, the behavior of dynamical SR is well-studied.
A mean-field solution of this model predicts a train of hyperbolic secant pulses for the number of photons in the cavity starting from a fully inverted initial state~\cite{bonifacio_coherent_1970}, and finite-size quantum corrections to this solution have been considered in Refs.~\cite{keeling_quantum_2009, babelon_semi-classical_2009}.
The same dynamics have also been studied in the context of cold gases~\cite{andreev_nonequilibrium_2004, barankov_atom-molecule_2004, szymanska_dynamics_2005, yuzbashyan_integrable_2005}, semiconductor microcavities~\cite{eastham_new_2007, eastham_quantum_2009}, impurity spins coupled to a microwave resonator~\cite{rose_coherent_2017},
and free-electron lasers~\cite{kling_high-gain_2021}.
There further exist studies of complementary initial states, where all emitters are in the ground state and the photon mode is in a Fock state ~\cite{strater_nonequilibrum_2012} or a coherent state~\cite{jurgens_comparison_2021}. Studies of the dynamics in low-excitation subspaces are also given in Refs.~\cite{tsyplyatyev_dynamics_2009, tsyplyatyev_classical_2010}.

Approximating physical emitters as ideal two-level systems is, however, not typically realistic; furthermore, additional degrees of freedom may lead to richer light-matter coupling phenomena.
For example, emitters in solid-state environments are subject to coupling to phonon modes and other sources of dephasing.
With organic molecules in particular, there are non-trivial effects of vibrational coupling to the electronic degrees of freedom~\cite{cwik_polariton_2014,herrera_theory_2018, wellnitzDisorderEnhancedVibrational2022},
unless one explicitly works to suppress vibrational excitations~\cite{wang_turning_2019}. 

Organic emitters have been used in experiments demonstrating strong light-matter coupling, polariton lasing, and Bose-Einstein condensation~\cite{torma_strong_2014,hakala_boseeinstein_2018, de_giorgi_interaction_2018,keeling_boseeinstein_2020, vakevainen_sub-picosecond_2020}. Hence, 
they are a promising platform for harnessing collective effects in light-matter coupling.
Motivated by the necessity to model realistic emitters for state-of-the-art experiments, we extend in this work the treatment of dynamical SR from two-level systems \cite{abate_exakte_1964,bonifacio_coherent_1970,barnett_collective_1984,seke_influence_1989,seke_collapse_1989,eastham_new_2007} to organic molecules, by explicitly incorporating vibrational modes in the system description.

As a first approximation, the effect of a vibrational environment on the electronic degrees of freedom can be treated as a pure dephasing term in a Markovian master equation, see Fig.~\ref{fig:system-initial-conditions}(a)i. This is a phenomenological model aiming to capture the decoherence induced by a vibrational bath that weakly couples to the electronic states, but crucially cannot account for realistic electron-phonon interactions. To go beyond the pure dephasing model, we include one vibrational mode per molecule explicitly in the system Hamiltonian, giving the Holstein--Tavis--Cummings (HTC) model~\cite{roden_accounting_2012, cwik_polariton_2014,spano_optical_2015, herrera_cavity-controlled_2016, wu_when_2016, zeb_exact_2018}, Fig.~\ref{fig:system-initial-conditions}(a)ii. The main advancement of this work is to study dynamical SR in the HTC model.
We will see that, while in both the phenomenological model and the HTC model vibrational coupling suppresses the extent and duration of the dynamical SR, 
there are significant differences in behavior, particularly when the vibronic coupling becomes strong.

The leading challenge of modelling dynamical SR systems is the exponential growth of the size of the Hilbert space with the number of molecules. For the open TC model with local incoherent processes, methods leveraging the weak permutation symmetry of the emitters~\cite{chaseCollectiveProcessesEnsemble2008,xu_simulating_2013, richter_numerically_2015, damanet_cooperative_2016, kirton_suppressing_2017, shammah_open_2018,barberenaGeneralizedHolsteinPrimakoffMapping2025} allow 
numerical solution up to $N\approx30$ emitters.
Here, we extend a powerful numerical method~\cite{gong_steady-state_2016}, which additionally exploits the weak U(1) symmetry of the Lindblad master equation, 
enabling a solution of the dissipative TC model with local losses and dephasing up to $N\sim140$ emitters when starting from a fully inverted initial state.

Experiments, however, have significantly larger numbers of emitters, and so other methods are required to reach large $N$.  
For this, we use a standard mean-field approach (assuming factorization of the molecular and cavity degrees of freedom), along with a second-order cumulant approach that captures leading-order corrections to the mean-field theory.
Such mean-field and cumulant approaches can generally be expected to match the exact solution in the thermodynamic limit $N\rightarrow\infty$ \cite{eastham_bose_2001, mori_exactness_2013, carollo_exactness_2021}, however, this is not always the case \cite{robicheaux_beyond_2021,rubies-bigorda_characterizing_2023,fowler-wright_determining_2023,stitely_quantum_2023,carollo_non-gaussian_2023}.
By comparing the exact solution to mean-field and second-order cumulant solutions, we show that the early-time behavior of dynamical SR in the TC model can be captured with both of these approximate methods if the process starts in an SR initial state ($\theta \neq 0$ in Fig.~\ref{fig:system-initial-conditions}(b)), not in the SF ($\theta = 0$) one. Since the second-order cumulant solution converges to the mean-field solution in the large $N$ limit, this motivates using mean-field theory to study dynamical SR with vibrationally dressed molecules in the HTC model.

Cumulant approaches have previously been used to study polariton lasing in the HTC model~\cite{arnardottir_multimode_2020, moilanen_mode_2022}. However, we note
that these works use a reduced set of 
cumulant equations where terms that vanish
due to U(1) symmetry are omitted.
Whilst appropriate for studying dynamics
from homogeneous initial conditions, such
an approach cannot be used to describe
SR initial conditions where there is an
initial breaking of this symmetry.

We point out another related work~\cite{fowler-wright_efficient_2022}
where mean-field theory was combined with the process-tensor formalism to solve the HTC model with a continuum of vibrational modes, treated as a non-Markovian bath. 
In the present work we model the situation
where there is one (or a few) vibrational
modes strongly coupled to the electronic
state, and show how higher-order correlations
are included via cumulants.


Our key findings are that strong vibronic coupling destroys dynamical SR, which is most clearly seen in the suppression of the buildup of electronic coherence. However, for weak and moderate vibronic coupling---closer to the typical values in organic molecules---SR is still possible. Interestingly, the vibrational coupling can even assist superradiance when the cavity frequency is smaller than the electronic transition frequency (negative cavity detuning). While a pure dephasing model can account for the suppression of SR, it fails to capture the dependence on the cavity detuning predicted in the HTC model, which is a measurable effect in experiments.


The structure of the paper is as follows. In Sec.~\ref{sec:pibs}, we introduce the dissipative TC model and our exact numerical solution method. We benchmark mean-field and second-order cumulant solutions against the exact solution for SF and SR initial conditions, and for different numbers of emitters.
In Sec.~\ref{sec:htc}, we introduce the open HTC model, compute its dynamics using mean-field theory, and compare it to the TC model with a pure dephasing term. Section \ref{sec:conclusion} concludes the paper.

\section{Numerical solution for the dissipative Tavis--Cummings model}
\label{sec:pibs}
\subsection{Model description}
We start by considering a system of $N$ two-level systems coupled to a single cavity mode, as in Fig~\ref{fig:system-initial-conditions}(a)i. In the dipole and rotating-wave approximations, such a system is well described by the TC Hamiltonian~\cite{tavis_exact_1968}.
Setting \(\hbar=1\), the system Hamiltonian is
\begin{equation}
    \opr{H}_\text{TC} = \wc \opr{a}^\dagger \opr{a} + \sum_{i=1}^N\biggl[\frac{\omega_0}{2}\opr{\sigma}_i^z + g(\opr{a}\opr{\sigma}_i^+ + \opr{a}^\dagger \opr{\sigma}_i^-)\biggr],\label{eq:H_tc}
\end{equation}
where $\opr{a}$ and $\opr{a}^\dagger$ are the annihilation and creation operators of the cavity mode of frequency $\wc$, $\opr{\sigma}_i^{z,\pm}$ are Pauli matrices describing the $i^\text{th}$ emitter satisfying $[\opr{\sigma}_i^z, \opr{\sigma}_j^\pm] =\pm2\delta_{ij}\opr{\sigma}_i^\pm$, $\omega_0$ is the energy splitting of the emitters, and $g$ is the individual light-matter coupling (which depends on the mode volume).
To include loss and dephasing processes, we employ a GKSL master equation for the density matrix $\rho$~\cite{breuer_theory_2002},
\begin{align}
      \partial_t\rho &= -\im[\opr{H}_\text{TC}, \rho] +\mathcal{D}_\text{TC}\label{eq:master_tc}\\
      \mathcal{D}_\text{TC}&=\kappa\lblad{\opr{a}}+\sum_{i=1}^N\bigl(\gamma\lblad{\opr{\sigma}_i^-} +\gamma_\phi\lblad{\opr{\sigma}_i^z}\bigr),
      \label{eq:D_tc}
\end{align}
with $\lblad{\opr{X}}= \opr{X}\rho \opr{X}^\dagger - \{\opr{X}^\dagger \opr{X},\rho\}/2$. 
The rate $\kappa$ describes the loss of photons from the cavity, $\gamma$ individual loss of molecular excitations (e.g., by spontaneous emission into a non-confined radiation mode), and $\gamma_\phi$ individual molecular dephasing. The last pure dephasing term can be 
understood as a Markovian approximation to the effect of vibrational modes coupling to the emitters.

To treat dynamical SR, we choose an initial state ${\rho(0)=\dyad{\psi(\theta)}}$ with
\begin{equation}
   \ket{\psi(\theta)}= \ket{0}\otimes\left(\bigotimes_{i=1}^N\ee^{-\im\theta\opr{\sigma}^x_i/2}\ket{\up}_i\right),\label{eq:tc_init}
\end{equation}
where $\ket{\up}_i$ is an excited (inverted) emitter state in the Hilbert space of the $i^\text{th}$ two-level system. The initial state has zero photons in the cavity mode, and the emitters are fully inverted and coherently tilted by an angle $\theta$. For $\theta=0$, the emitters are fully inverted, giving SF, whereas for $\theta\neq 0$, there is some initial dipole moment present, giving SR [Fig.~\ref{fig:system-initial-conditions}(b)]. 
Note that a rotation of the inverted states about any axis in the $xy$ plane (i.e. $\exp[-\im(n_x\opr\sigma_i^x + n_y\opr\sigma_i^y)\theta/2]$; $n_x^2+n_y^2=1$) leads to an emitter state with equivalent initial coherence as in Eq.~\eqref{eq:tc_init}, and we choose here the $x$-axis for simplicity (i.e. $n_x=1,\,n_y=0$). Such states of two-level systems are the analog of coherent states in bosonic systems, which are also referred to as spin-coherent states or atomic-coherent states~\cite{radcliffe_properties_1971, mandel_optical_1995}, and they have recently been used to study entanglement in Dicke SR \cite{rosarioUnravelingDickeSuperradiant2025}.


\subsection{Numerical method}
To solve Eq.~\eqref{eq:master_tc} numerically, we first make use of the weak permutation symmetry of the emitters~\cite{chaseCollectiveProcessesEnsemble2008,xu_simulating_2013, richter_numerically_2015, damanet_cooperative_2016, kirton_suppressing_2017, shammah_open_2018,barberenaGeneralizedHolsteinPrimakoffMapping2025} by expanding the density matrix as
\begin{equation}
    \rho = \sum_{\lambda,n,n'}\rho_{\lambda,n,n'}\dyad{n}{n'}\otimes \opr{O}_\lambda,
\end{equation}
where $\dyad{n}{n'}$ is the state of the cavity mode with photon numbers $n$ and $n'$, and $\opr{O}_\lambda$ is a permutation-symmetric density matrix depending on the $N$-emitter state labelled by $\lambda$. Since we consider two-level systems, the permutation-symmetric states can be labelled by enumerating the ways of dividing $N$ elements into four bins corresponding to the possible emitter matrix elements of a single two-level system $\up\up,\down\up,\up\down$, and $\down\down$ ($\uparrow$: excited state, $\downarrow$: ground state). Thus, $\lambda$ describes the $4$-element lists  $\vb{m}_\lambda = (m_{\lambda,\up\up},m_{\lambda,\down\up},m_{\lambda,\up\down},m_{\lambda,\down\down})$ with $m_{\lambda,p}\geq 0$ and $\sum_p m_{\lambda,p}=N$.
Explicitly, each $\vb{m}_\lambda$ corresponds to a matrix $\opr{O}_\lambda$ that consists of the sum of all states obtainable by permuting pairs of emitters in a state with $m_{\lambda,\up\up}$ inverted emitters in the left and right states of the density matrix, followed by $m_{\lambda,\down\up}$ ground-state emitters in the left state and inverted emitters in the right state, and so on. 

Using the mapping described above, one can, in principle, construct the operators $\opr{O}_\lambda$ in the original explicit $2^N$ dimensional two-level-system basis.  The permutation approach is, however, based on avoiding the explicit representation and performing all calculations within the compressed space.
The number of unique density-matrix elements in the emitter space is given by the number of possible partitions $\vb{m}_\lambda$, which is equal to $\binom{N+3}{3}$.

Since the initial state in Eq.~\eqref{eq:tc_init} has $N$ excitations in the emitter state and there are no gain terms in Eq.~\eqref{eq:master_tc}, the photon Hilbert space can be truncated at $N+1$ without introducing any approximations. 
The total number of density-matrix elements therefore scales as $(N+1)^2\times \binom{N+3}{3}$, and an exact solution of Eq.~\eqref{eq:master_tc} becomes possible up to order of $N\sim30$ emitters.

We can further exploit the weak~\cite{buca_note_2012,albert_symmetries_2014} U(1) symmetry of the master equation \eqref{eq:master_tc}, which states that it is invariant under $\opr{a}\rightarrow \opr{a} \ee^{\im\phi},\,\opr{\sigma}_n^-\rightarrow \opr{\sigma}_n^-\ee^{\im\phi}$. 
Let us denote the total number of excitations in the left and right states of a given density-matrix element $\rho_{\lambda,n,n'}$ as $\nu=n+m_{\lambda,\up\up}+m_{\lambda,\up\down}$ and $\nu'=n'+ m_{\lambda,\up\up} + m_{\lambda,\down\up}$. As a consequence of the weak U(1) symmetry, only elements with a constant value of $\nu-\nu'$ can couple in the dynamics. This motivates the introduction of the block form of the density matrix
\begin{equation}
        \rho = \bigoplus_{\nu,\nu'}\rho^{(\nu,\nu')},
\end{equation}
where $\rho^{(\nu,\nu')}$ denotes the collection of all elements with left and right excitation numbers equal to $\nu$ and $\nu'$, respectively. 
The most general equations of motion for the individual blocks respecting the weak U(1) symmetry then read
\begin{equation}
\partial_t{\rho}^{(\nu,\nu')}=\sum_{k\in\mathbb{Z}}L_k^{(\nu,\nu')}\rho^{(\nu+k,\nu'+k)}, \label{eq:supercomp_general}
\end{equation}
where $L_k^{(\nu,\nu')}$ are matrices determined by the master equation.
All terms with $k<0$ describe gain processes, $k=0$ describes the coherent dynamics and dephasing, and $k>0$ describes losses. Note that blocks with different values of $\nu$ and $\nu'$ do couple, but their difference $\nu-\nu'$ is constant.
Such a structure of master equation---combining weak U(1) symmetry with weak permutation symmetry---was used in Ref.~\cite{gong_steady-state_2016} to discuss steady state superradiance in a system of Rydberg atoms.

Since Eq.~\eqref{eq:master_tc} features jump operators that change the total excitation number by at most one, and the system we consider has no gain, only the terms $k=0$ and $k=1$ are non-zero in Eq.~\eqref{eq:supercomp_general}. Further, we are primarily
interested in calculating the expectation value of the number of photons in the cavity mode $\langle \opr{n}\rangle = \langle \opr{a}^\dagger \opr{a}\rangle$, so we can restrict Eq.~\eqref{eq:supercomp_general} to only those blocks with $\nu-\nu'=0$
on which the photon number operator depends. 
In this case, we arrive at the evolution equation for the blocks $\rho^{(\nu)}\equiv\rho^{(\nu,\nu)}$
\begin{equation}
\partial_t\rho^{(\nu)}=L_0^{(\nu)}\rho^{(\nu)} + L_1^{(\nu)}\rho^{(\nu+1)}. \label{eq:master_block}  
\end{equation}
The matrices $L_0^{(\nu)}$ and $L_1^{(\nu)}$ can be constructed from Eq.~\eqref{eq:master_tc}.
Crucially, since we consider a system without gain, we can solve Eq.~\eqref{eq:master_block} sequentially by noting that there exists a block with the highest excitation number $\nu_\text{max}$. For this block, the second term on the right-hand side in Eq.~\eqref{eq:master_block} vanishes, resulting in a homogeneous system of differential equations for the block $\rho^{(\nu_\text{max})}$, which can be numerically integrated. 
The solution $\rho^{(\nu_\text{max})}$
may then be used to solve 
the block $\rho^{(\nu_\text{max}-1)}$.
Iterating until $\nu=0$ yields dynamics for all blocks with \(\nu-\nu'=0\).
We note that this method of solution is reminiscent of that used to solve
the diagonal elements of the density matrix in Dicke SR~\cite{bonifacioQuantumStatisticalTheory1971}.


From the initial state in Eq.~\eqref{eq:tc_init}, we see that
the maximum excitation number is $\nu_\text{max}=N$.
The number of elements of the blocks $\rho^{(\nu)}$ for a fixed $N$ increases monotonically from $1$ at $\nu=0$
to $\binom{N+3}{3}$ at $\nu=N$. The largest block must have exactly as many elements as there are emitter states, because for every emitter state, one can always find photon numbers $n$ and $n'$ such that $\nu=\nu'=N$. 

Code for the numerical scheme described is available in the \emph{PIBS} (Permutational Invariance Block Solver) Python package~\cite{freterLukasfreterPibsVersion2025}.
This method allows one to solve Eq.~\eqref{eq:master_tc} with initial conditions given in Eq.~\eqref{eq:tc_init} up to $N=140$ emitters.
Currently, only the solution for the diagonal block $\nu=\nu'$ is implemented in the \emph{PIBS} code, but an extension to off-diagonal blocks is straightforward.


\subsection{Benchmarking Cumulant approaches}
\begin{figure}
  \centering
  \includegraphics[width=\linewidth]{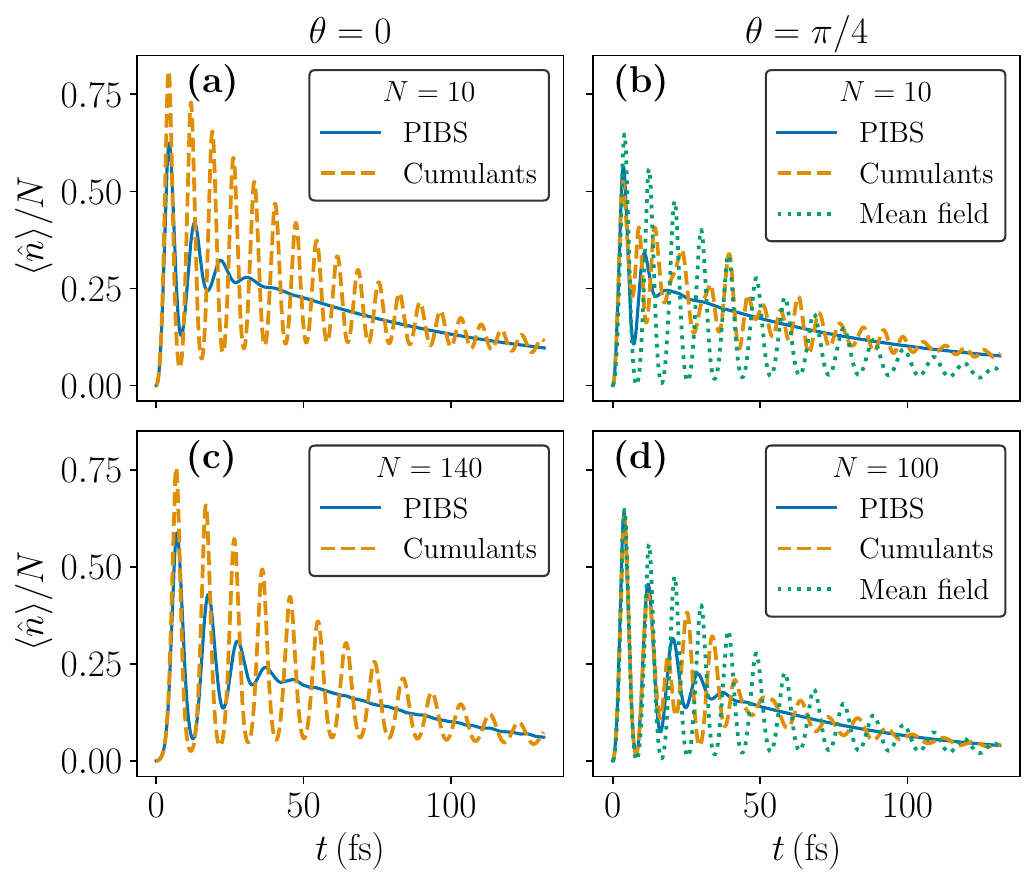}
  \caption{Comparison of exact dynamics of the average number of photons in the cavity calculated with \emph{PIBS} and cumulant approaches. Panels (a) and (c) show the dynamics for SF initial conditions for $N=10$ and $N=140$. Panels (b) and (d) show the dynamics for SR initial conditions ($\theta=\pi/4$) for $N=10$ and $N=100$. Other parameters: $g\sqrt{N}=0.4\eV,\,\Delta=\wc-\omega_0=-0.35\eV,\,\kappa=0.01\eV,\,\gamma=0.001\eV,\,\gamma_\phi=0.0075\eV$. }
  \label{fig:pibs_c2}
\end{figure}

In this section, we compare the exact solution of Eq.~\eqref{eq:master_tc} based on \emph{PIBS} to approximate first- and second-order cumulant approaches. In a cumulant approximation of order $M$, one splits expectation values of $M+1$ operators in the Heisenberg equations of motion into expectation values of at most $M$ operators by setting the cumulants of order $M+1$ to zero. For example, in a first-order cumulant approximation, the expectation values of any two operators $\opr{A}$ and $\opr{B}$ are split as $\langle \opr{A}\opr{B}\rangle = \langle \opr{A}\rangle \langle \opr{B}\rangle$, because the second-order cumulant given by $\langle \langle \opr{A}\opr{B}\rangle\rangle=\langle \opr{A}\opr{B}\rangle-\langle \opr{A}\rangle\langle \opr{B}\rangle$~\cite{gardinerStochasticMethodsHandbook2009} is set to zero. This is equivalent to mean-field theory. Going one order higher, keeping expectations of pairs of operators and setting third-order cumulants to zero, one arrives at the second-order cumulant approximation.

If the initial state respects the U(1) symmetry of the Hamiltonian in Eq.~\eqref{eq:H_tc}, one can simplify the second-order cumulant equations by dropping expectation values of operators that do not conserve the number of excitations, such as $\langle \opr{a}\rangle$ and $\langle\opr{\sigma}_i^+\rangle$. If these ``symmetry-breaking'' terms are zero initially, 
then they remain zero for all times. The resulting equations may be
called symmetry-preserving cumulant equations~\cite{fowler-wright_determining_2023}.

The initial state in Eq.~\eqref{eq:tc_init} breaks the U(1) symmetry for $\theta\neq 0$, since $\langle\opr{\sigma}_i^+\rangle = \im\sin(\theta)/2$. As a consequence, for SR initial conditions, one must retain
all terms in the cumulant expansion, resulting in symmetry-broken cumulant equations. We use the \emph{QuantumCumulants.jl} 
\emph{Julia} package~\cite{plankensteiner_quantumcumulantsjl_2022} to calculate 
and evolve the symmetry-broken cumulant equations.

In Fig.~\ref{fig:pibs_c2}, we compare the cumulant approaches to the exact solution for different numbers of emitters and initial conditions, by plotting the normalised average number of photons in the cavity $\langle \opr{n}\rangle/N=\langle \opr{a}^\dagger \opr{a}\rangle/N$ as a function of time. For brevity, we refer to the first-order cumulant approximation as mean field and to the second-order cumulant approximation simply as cumulants. We do not show a mean-field solution for SF initial conditions, as the mean-field solution in that case is just zero for all times if we start from the symmetric initial state. We also note that in the mean-field treatment, $\langle \opr{n}\rangle/N=|\langle\opr{a}\rangle|^2/N$ becomes independent of $N$.

Figures~\ref{fig:pibs_c2}(a) and \ref{fig:pibs_c2}(c) show the dynamics for $N=10$ and $N=140$ for SF initial conditions ($\theta=0$). One can see that cumulants underestimate the dephasing processes, as the amplitude of oscillations is much larger than that in the exact solution. 
As such, for $N=140$, we only see agreement between the cumulant approximation and the exact solution at early times, specifically for 
$t<10\fs$.  At later times, there is no agreement over the range of $N$ for which we can obtain exact results.
Part of the dephasing (which is different from the dephasing due to the vibrations) causing this mismatch can be explained by the quantum corrections to the energy eigenvalues of the TC Hamiltonian for finite $N$~\cite{keeling_quantum_2009}. 
As discussed there, the mean-field (classical) solution with periodic oscillations corresponds to the case of equal (harmonic) energy-level spacing.  
At finite $N$ however the level spacing is not perfectly regular.
Using the fact that the resulting anharmonicity is known to scale as $1/(\ln N)^3$, and that the cumulant solution for $N=140$ matches the exact solution up to $t\approx10\fs$ [cf. Fig.~\ref{fig:pibs_c2}(c)], we can estimate that, e.g., for $N=10^8$, 
this dephasing would set in after $t\approx 520\fs$.

Figures~\ref{fig:pibs_c2}(b) and \ref{fig:pibs_c2}(d) compare the exact solution to cumulants and mean field for an SR initial state with $\theta=\pi/4$. Generally, the cumulant solution matches the exact solution better than the mean-field solution. Moreover, for increasing $N$, the early time period where cumulants match the exact solution increases. This indicates that for SR initial states, symmetry-broken cumulants capture the early-time behavior 
well. We provide quantitative evidence
that this is the case in Appendix~\ref{sec:appendix_convergence},
where we show
that both mean-field and cumulants converge monotonically towards the exact solution for $N$ up to $100$.

\begin{figure}
  \centering
  \includegraphics[width=\linewidth]{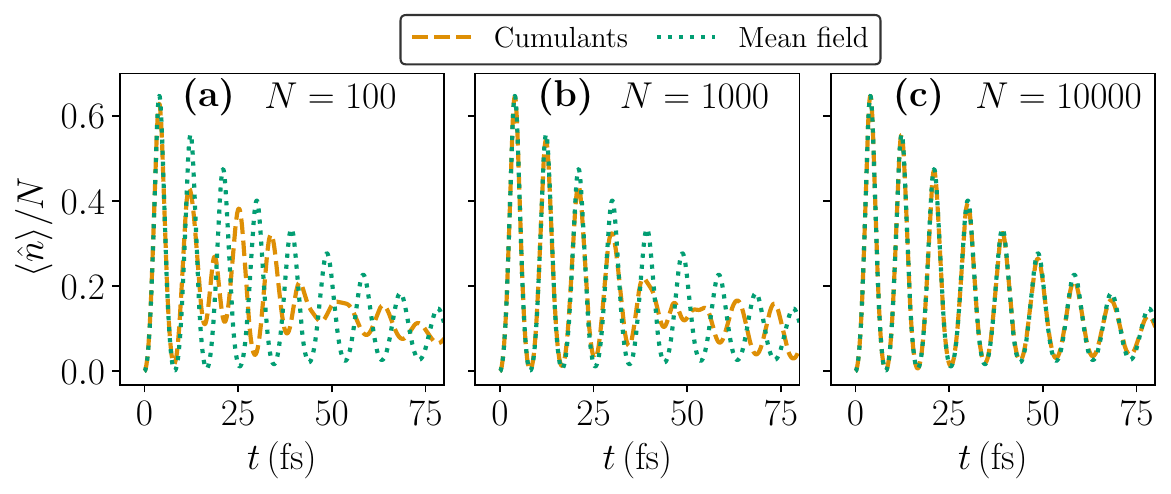}
  \caption{Comparison of symmetry-broken second-order cumulant approximation to mean-field approximation for (a) $N=100$ [same as Fig.~\ref{fig:pibs_c2}(d)], (b) $N=1000$, and (c) $N=10000$. The SR angle is set to $\theta=\pi/4$, all other parameters are the same as in Fig.~\ref{fig:pibs_c2}.}
  \label{fig:tc_c2_mf}
\end{figure}

As an additional test of mean-field results, we note that for SR initial states, the symmetry-broken cumulant solution converges to the mean-field solution for large enough $N$, as demonstrated in Fig.~\ref{fig:tc_c2_mf}. Since we are ultimately interested in the SR initial state, this leads us to conclude that for large $N$, it is reasonable to use mean-field theory.

\section{Superradiance with coupling to vibrations}
\label{sec:htc}
\subsection{Model description}
In the previous section, we discussed the dissipative TC model with local dephasing, capturing 
the emitters vibronic coupling within a Markovian approximation.
We now introduce a model that includes vibrations explicitly in the system dynamics.
As a simplest case, we assume that there is one dominant vibrational mode of frequency $\omega_\nu$, which is the same for all molecules. 
We extend the TC model from Eq.~\eqref{eq:H_tc} to include this mode, 
\begin{equation}
    \opr{H}_\text{HTC} = \opr{H}_\text{TC} + \sum_{i=1}^N\omega_\nu\bigg[ \opr{b}^\dagger_i \opr{b}_i+\sqrt{S}(\opr{b}_i+\opr{b}_i^\dagger)\opr{\sigma}_i^z \biggr],\label{eq:h_htc}
\end{equation}
which is the HTC model 
\cite{roden_accounting_2012, cwik_polariton_2014}.

The vibrational mode of the $i^\text{th}$ molecule is treated as a harmonic oscillator with annihilation and creation operators $\opr{b}_i$ and $\opr{b}_i^\dagger$. The coupling 
strength between the vibrations and the electronic degrees of freedom is set
by the Huang-Rhys parameter $S$. 
Including the same dissipation processes as before (see Eq.~\eqref{eq:D_tc}~\footnote{We note that since the dominant vibrational mode is included in the system description in the HTC model, the dephasing terms in Eq.~\eqref{eq:master_htc} and Eq.~\eqref{eq:master_tc} have different meanings in principle. However, since we only set $\gamma_\phi>0$ in case $S=0$, the two dephasing rates are equivalent.}), and adding thermalization processes for the vibrational modes, the master equation is  
\begin{multline}
      \partial_t\rho=-\im[\opr{H}_\text{HTC}+\opr{H}_\text{LS},\rho] + \mathcal{D}_\text{TC}  \\+\sum_{i=1}^N\bigl(\gamma_\up\lblad{\opr{b}_i^\dagger}+\gamma_\down\lblad{\opr{b}_i}\bigr),\label{eq:master_htc}
\end{multline}
where $\mathcal{D}_\text{TC}$ was defined in Eq.~\eqref{eq:D_tc}.
The thermalization rates are $\gamma_\up = \gamma_\nu n_\text{B}$ and $\gamma_\down = \gamma_\nu(n_\text{B}+1)$, where $n_\text{B} = (\ee^{\omega_\nu/T}-1)^{-1}$ is the Bose-Einstein distribution at temperature $T$ (we set $k_\text{B}=1$).
The model of thermalization here,
which includes a 
Lamb-shift term $\opr{H}_\text{LS} = \im(\gamma_\nu/4)\sum_{i=1}^N(\opr{b}_i^\dagger \opr{b}_i^\dagger-\opr{b}_i\opr{b}_i)$, is one of momentum damping,
as discussed in Ref.~\cite{mannouch_ultra-fast_2018}.
The energy of the HTC Hamiltonian~\eqref{eq:h_htc} depends on the vibrational displacement $\opr{x}_i\propto(\opr{b}_i^\dagger+\opr{b}_i)$, and the thermalization drives the system towards $\opr x_i$ such that the energy is minimized.
In Figure~\ref{fig:system-initial-conditions}(a)ii, the internal level structure and incoherent processes of the HTC model are shown.

Using Eq.~\eqref{eq:tc_init}, we can write the initial state including vibrations as
\begin{equation}
    \rho(0) = \dyad{\psi(\theta)}\bigotimes_{i=1}^N\rho_{\text{vib},i},\label{eq:htc_init}
\end{equation}
where $\rho_{\text{vib},i} = (1-\ee^{-\omega_\nu/T})\ee^{-\omega_\nu\opr{b}^\dagger_i\opr{b}_i/T}$ is a thermal state of the vibrational mode of molecule $i$. 

The Hilbert space of the HTC model is much larger than the one considered in Sec.~\ref{sec:pibs}. Truncating the vibrational mode to $N_\nu$ levels increases the Hilbert space by a factor of $N_\nu^N$, which makes an exact solution quickly intractable for growing $N$. However, from Sec.~\ref{sec:pibs} we have confidence in mean-field theory 
capturing the essential early-time behavior also for the HTC model, and therefore solve the dynamics using the mean-field approximation.

In what follows, we compare the dynamics of the model Eq.~\eqref{eq:master_htc} for two different cases [cf. Fig.~\ref{fig:system-initial-conditions}(a)]: (i) $S=0$ and variable $\gamma_\phi$, which just reduces to the TC model Eqs.~\eqref{eq:H_tc} and \eqref{eq:master_tc}, and (ii) $\gamma_\phi=0$ and variable $S$. We do this comparison because both $S$ and $\gamma_\phi$ model forms of dephasing by phonons, and we are interested in how these different processes in solid-state environments modify superradiance.

\subsection{Results}
The results shown here were calculated in the mean-field approximation using the \emph{Julia} package \emph{QuantumCumulants.jl}~\cite{plankensteiner_quantumcumulantsjl_2022}.  We consider SR initial conditions with $\theta=10^{-3}\pi$ throughout, and set the number of molecules $N=10^8$.

\begin{figure*}
      \centering
      \includegraphics[width=1\linewidth]{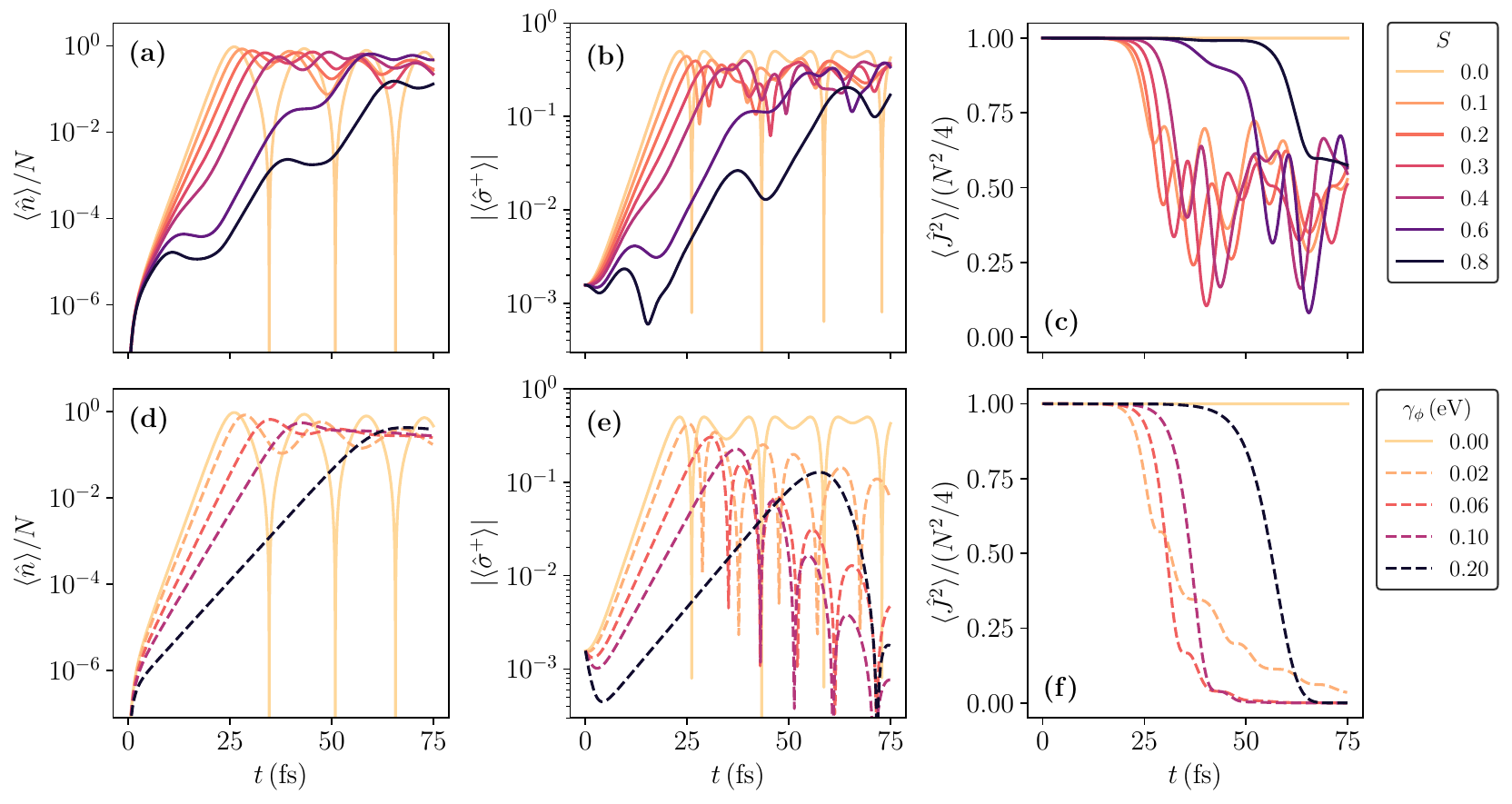}
      \caption{(a), (b) and (c) Average number of photons in the cavity mode, electronic coherence, and magnitude of the Bloch vector, respectively, as a function of time for different coupling strengths to the vibrational mode. (c), (d), (f) Same as (a), (b), (c) but with the vibrational mode replaced by a pure dephasing term with rate $\gamma_\phi$. The curves $S=0$ in (a), (b), (c) are the same as the curves with $\gamma_\phi=0$ in (d), (e), (f). 
      Note that the values of $S$ and $\gamma_\phi$ are not equidistantly spaced. Parameters:   $N=10^8,\,g\sqrt{N}=0.2\eV,\,\Delta=0,\,\kappa=0.01\eV,\,\gamma=10^{-6}\eV,\,\theta=10^{-3}\pi, \,\omega_\nu=0.15\eV,\,\gamma_\nu=0.01\eV,\,T=0.026\eV$. For (a), (b), (c): $\gamma_\phi=0$, for (d), (e), (f): $S=0$.}
      \label{fig:htc_phot}
\end{figure*}

Figure \ref{fig:htc_phot}(a) shows the average number of photons in the cavity mode as a function of time for different values of $S$. For $S=0$, the HTC model reduces to the TC model, resulting in damped Rabi-oscillations (note the logarithmic scale). For $S\leq0.3$, after a build-up time where $\langle\opr{n}\rangle/N$ grows from $0$ up to roughly $10^{-6}$, there is a region where the photon number increases exponentially with time as $\langle \opr{n}\rangle\propto \ee^{t/\tau}$. The parameter $\tau$ is the characteristic risetime, which is seen to increase with $S$. Increasing $S$ beyond a value of $0.3$ changes the exponential behavior after the build-up time, and a characterization based on a single risetime $\tau$ is no longer meaningful. It is, however, clear that a stronger coupling to the vibrational modes inhibits the emission of photons into the cavity mode.

In Fig.~\ref{fig:htc_phot}(d), the vibrational mode has been replaced by a pure Markovian dephasing term with rate $\gamma_\phi$. 
For all values of $\gamma_\phi$, there is a region where the photon number increases exponentially, and the risetime increases with increasing $\gamma_\phi$. 
Thus, the pure dephasing model can capture the photon dynamics for small $S$ below $0.3$, 
but breaks down for larger values of $S$. 

To characterize the extent to which the vibrational modes destroy superradiance, we additionally calculate the electronic coherence $\vert\langle\opr{\sigma}^+\rangle\vert\equiv\vert\langle\opr{\sigma}_i^+\rangle\vert$ for the same sets of parameters used in Figs.~\ref{fig:htc_phot}(a) and \ref{fig:htc_phot}(d), see Figs.~\ref{fig:htc_phot}(b) and \ref{fig:htc_phot}(e). 
Since we use SR initial conditions, the coherence at $t=0$ is non-zero and given by $\vert\langle\opr{\sigma}^+\rangle\vert = \sin(\theta)/2$. Figure \ref{fig:htc_phot}(b) shows that the coherence increases exponentially at early times for small $S$, similar to the exponential rise in the photon number in Fig.~\ref{fig:htc_phot}(a). For $S\gtrsim0.3$, the exponential increase of the coherence is inhibited, and there are regions where the coherence decreases before it has reached its maximum. In Appendix~\ref{sec:appendix}, we show that increasing the initial amount of coherence (increasing $\theta$) tends to make the initial rise of photon number faster and less dependent on the vibrational coupling, and in Appendix~\ref{sec:appendix_gSqrtN} we show that the threshold of SR at a certain value of $S$ (here at $S\approx0.3$) emerges as a competition between the light-matter coupling $g\sqrt{N}$ and the vibronic coupling $\omega_\nu\sqrt{S}$.

In Fig.~\ref{fig:htc_phot}(e), where the vibrational mode has been replaced by a pure dephasing term, we see an exponential increase of the coherence for low dephasing rates and early times, similar to the behavior in Fig.~\ref{fig:htc_phot}(b) for small $S$. Large dephasing rates also destroy superradiance, as can be seen by the initial decrease of coherence for $\gamma_\phi\gtrsim0.06\eV$.
We note that the long-time behavior of the coherence is quite different in Figs.~\ref{fig:htc_phot}(b) and \ref{fig:htc_phot}(e). In the former case, the coherence builds up eventually and remains relatively large, whereas in the latter case, the coherence decreases. This shows that, when regarding $\langle\sigma^+\rangle$, the coupling to vibrations does not result in the same kind of decoherence as introduced by a pure dephasing term, even for small $S$.

Lastly, Figs.~\ref{fig:htc_phot}(c) and~\ref{fig:htc_phot}(f) show the magnitude of the Bloch vector 
$ \langle\opr{J}^2\rangle = \langle\opr{J}_x^2+\opr{J}_y^2+\opr{J}_z^2\rangle$, $\opr{J}_\alpha=\sum_{i=1}^N\opr\sigma_i^\alpha/2$, which, within mean-field theory and using $N\gg1$, is given by 
\begin{equation}
    \langle\opr{J}^2\rangle = \frac{N^2}{4}\left(\langle\opr\sigma^z\rangle^2+4\vert\langle\opr{\sigma}^+\rangle\vert^2\right).\label{eq:bloch_mf}
\end{equation}
The initial state in Eq.~\eqref{eq:tc_init} is an eigenstate of $\opr{J}^2$ with the maximum possible eigenvalue $N/2(N/2+1)\approx N^2/4$.
In the absence of incoherent processes and vibronic coupling, i.e. for the closed TC model in Eq.~\eqref{eq:H_tc}, the emitter state remains an eigenstate of $\opr{J}^2$ with the same eigenvalue since $[\opr{H}_\text{TC},\opr{J}^2]=0$, and therefore $\langle \opr{J}^2\rangle\approx N^2/4$ for all times.
This is seen for the curves with $S=0$ or $\gamma_\phi=0$ in Figs.~\ref{fig:htc_phot}(c) and~\ref{fig:htc_phot}(f). We note that the effect of local molecular losses proportional to $\gamma=10^{-6}\eV$ is negligible in the time period considered here, which would otherwise lead to a decrease in $\langle \opr{J}^2\rangle$.

For non-zero $S$ or $\gamma_\phi$, the magnitude of the Bloch vector does not retain its maximum initial value. In the case of vibrational dressing in Fig.~\ref{fig:htc_phot}(c), $\langle \opr{J}^2\rangle$ is subject to oscillations, but stays well above zero in the considered time frame. In contrast, for pure dephasing in Fig.~\ref{fig:htc_phot}(f), $\langle \opr{J}^2\rangle$ decreases monotonically to zero, with minor bumps for small dephasing rates, showing a stark difference between the two models of treating the vibrational effects.

Such behavior can be related to discussions of how any processes that break spin conservation destroy free-space superradiance~\cite{friedbergLimitedSuperradiantDamping1972,
friedberg_temporal_1974,coffey_effect_1978}.  
As noted in those works, when the dynamics approaches the equator---specifically when $|\langle \hat\sigma^z \rangle| \lesssim 1/\sqrt{N}$---simple counting arguments show there are many more accessible subradiant states (quantum states with $\langle \hat J^2 \rangle \ll N$) than superradiant states.  As such, any process that can change the modulus of $\hat J$ will lead to transitions to these subradiant states, reducing $\langle \hat J^2 \rangle$ and ultimately destroying the superradiance.
Notably, these processes only become relevant as one approaches the equator:  when $\langle\hat \sigma^z \rangle \simeq 1$, scattering to subradiant states is suppressed if the scattering does not directly change  $\langle \hat\sigma^z \rangle$.
References~\cite{friedbergLimitedSuperradiantDamping1972,
friedberg_temporal_1974,coffey_effect_1978} discuss how direct interactions between different emitters, breaking the permutation symmetry, have such an effect.  Here we see similar behavior driven by individual dephasing or vibrational modes.

Interestingly, for both models, a stronger influence of the vibrational modes (i.e., larger $S$ or larger $\gamma_\phi$) leads to an increased time period, for which $\langle \opr{J}^2\rangle$ stays close to its maximum initial value. This can be explained by the decreased emission rate of photons for larger $S$ and $\gamma_\phi$, see Figs.~\ref{fig:htc_phot}(a) and~\ref{fig:htc_phot}(d), which means that the excitation state remains close to its initial value of $\langle\opr{\sigma}^z\rangle\approx 1$ for longer times, and thus avoids states near the equator where transitions to subradiant states arise. According to Eq.~\eqref{eq:bloch_mf}, this results in $\langle\opr{J}^2\rangle\approx N^2/4$.

\begin{figure}
  \centering
  \includegraphics[width=1\linewidth]{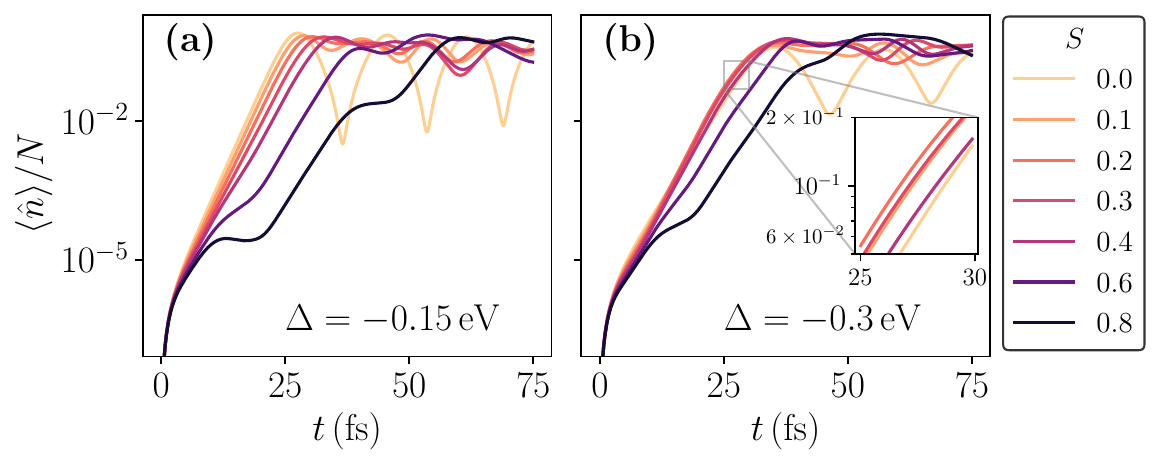}
  \caption{Average number of photons in the cavity mode as a function of time for different coupling strengths to the vibrational mode for (a) $\Delta=-0.15\eV$ and (b) $\Delta = -0.3\eV$. All other parameters are the same as in Fig.~\ref{fig:htc_phot}, and we specifically note that the vibrational frequency is set to $\omega_\nu=0.15\eV$.}
  \label{fig:htc_detuning}
\end{figure}


\begin{figure}
  \centering
  \includegraphics[width=1\linewidth]{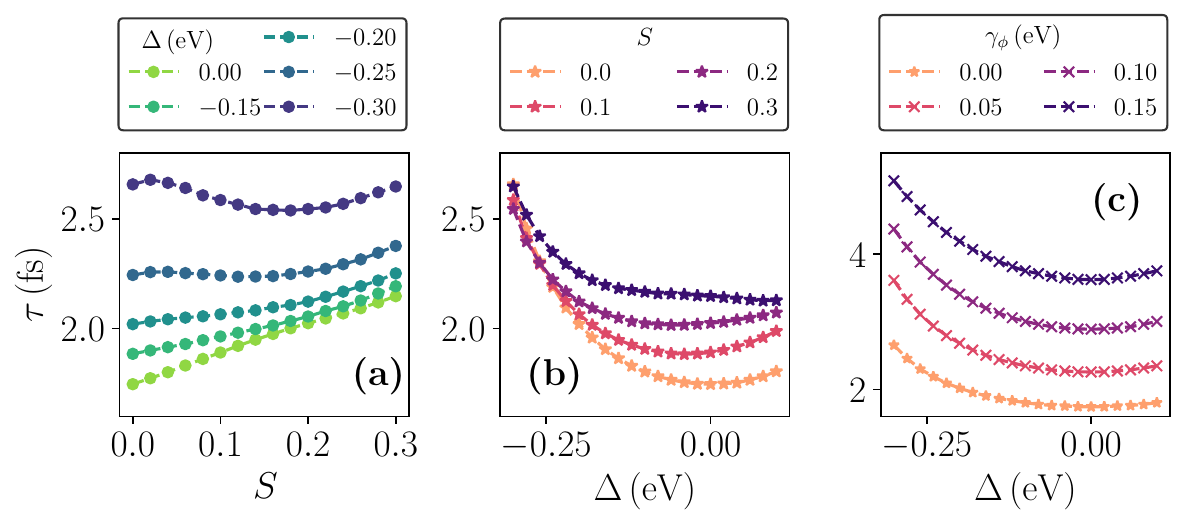}
  \caption{Photon number risetime extracted from an exponential fit to the linear regions as seen for example in Fig.~\ref{fig:htc_phot}(a) and (c). (a) Risetime as a function of $S$ for different detunings. (b) Risetime as a function of detuning for different values of $S$. (c) Same as (b), but the vibrational mode has been replaced by a pure dephasing term with rate $\gamma_\phi$. Note that the curve with $S=0$ in (b) is the same as the curve with $\gamma_\phi=0$ in (c). All other parameters are the same as in Fig.~\ref{fig:htc_phot}. }
  \label{fig:risetime}
\end{figure}

In experiments, the cavity frequency can typically be readily varied; our calculations predict interesting phenomena when varying the cavity detuning $\Delta = \wc - \omega_0$, with a clear difference between pure dephasing and having coupling to a vibrational mode.
The dynamics of the photon mode in the HTC model is plotted for $\Delta=-0.15\eV$ in Fig.~\ref{fig:htc_detuning}(a) and for $\Delta = -0.3\eV$ in Fig.~\ref{fig:htc_detuning}(b).
Compared to the case of zero detuning in Fig.~\ref{fig:htc_phot}(a), we see that a negative detuning favors the emission of photons for larger values of $S$. This is seen most clearly in Fig.~\ref{fig:htc_detuning}(b), where the curves up to $S=0.4$ almost collapse into one curve at early times. In the inset of Fig.~\ref{fig:htc_detuning}(b), we 
observe that the photon number rises faster for $0.1\leq S\leq0.4$ than for $S=0$. This can be explained by first noting that, for $\omega_\nu=0.15\eV$ and $T=0.026\eV$ used here, the phonons are almost fully in the ground state at $t=0$. Further, for a negative detuning, the cavity frequency $\wc = \omega_0+\Delta$ is smaller than the level splitting of the electronic ground and excited $\omega_0$. Thus, the cavity frequency matches transitions from the excited state manifold with zero vibrational excitations to the ground-state manifold with vibrational excitations. 
As $S$ increases, the most probable number of vibrational excitations created in an optical transition increases.  
As such, one sees an optimal value of $S$ in this regime where processes exciting one vibrational mode become dominant. In contrast, if the vibrational mode is treated as a pure dephasing term, such a reordering phenomenon does not occur.

For a final set of results, we calculate the risetime $\tau$ for different values of $S$, $\Delta$, and $\gamma_\phi$, while making sure to stay in a region where $\tau$ is well defined. Figure~\ref{fig:risetime}(a) shows $\tau$ as a function of $S$ and different values of $\Delta$. For small negative detunings, $\tau$ increases monotonically with $S$, as evident in Fig.~\ref{fig:htc_phot}(a). In contrast, large negative detunings $\Delta\leq-0.25\eV$ show regions where increasing $S$ leads to a shorter risetime, which is the behavior observed in Fig.~\ref{fig:htc_detuning}. This behavior is confirmed in Fig.~\ref{fig:risetime}(b), which shows $\tau$ as a function of $\Delta$ for different $S$. Here we see in particular how the minimum of $\tau$ at zero detuning for the TC model ($S=0$) shifts to more negative detunings as $S>0$ is increased, and further, the symmetry about the minimum value of $\tau$ is lost. This symmetry remains in the case where the vibrational mode is replaced by a pure dephasing rate, irrespective of the strength of the dephasing, see Fig.~\ref{fig:risetime}(c). In that case, the risetime increases as $\gamma_\phi$ increases, and the minimum value is always at zero detuning.

The risetime behavior for different values of $\theta$ is discussed in Appendix \ref{sec:appendix}. We find that there is an initial rapid rise to a higher photon number for large $\theta$ than for a small one. That is, the photon number rise is faster for a smaller amount of electronic excitation (smaller inversion). This is a curious behavior, opposite, for example, to usual pulsed lasing, where the pulse build-up time becomes shorter for a higher amount of excitation~\cite{vakevainen_sub-picosecond_2020}. This is because coherence plays a crucial role in SR, and larger $\theta$ (up to $\pi/2$) means higher initial coherence.  


\section{Conclusions}
\label{sec:conclusion}
We have studied how processes of vibrational coupling and pure
dephasing---relevant to realistic models of emitters in solid-state
environments---affect the dynamics of superradiance and superfluorescence for
organic molecules in an optical cavity.  To enable this study, we
introduced an efficient method leveraging both weak permutation and weak U(1)
symmetry of the many-molecule--cavity system that allows an exact solution for a large
number of emitters.  For the Tavis--Cummings model with dissipation and
dephasing, solutions were obtained up to \(N\sim140\) emitters.  While we
focused the exact method
on the simple case of two-level emitters in a single-mode cavity, it can be extended to handle more
complex multi-level systems and other near-resonant cavity modes.
This offers a valuable tool to benchmark other numerically exact and
approximate methods for cavity-QED systems, greatly extending the range of
system sizes that can be handled exactly. 

Using our exact method, we benchmarked mean-field and second-order cumulant
solutions, verifying their accuracy in capturing early-time dynamics when the
initial state has non-zero initial coherence. This justified the application of
mean-field theory to study the effect of the vibrational environment of organic
emitters on the dynamics. Here we compared two models: a Holstein--Tavis--Cummings
model including a discrete vibrational mode for each emitter, and a
simplified phenomenological model where the vibronic coupling is replaced by Markovian pure dephasing. We
found that for small vibrational coupling or dephasing, both models predict an exponential
rise of the photon number at early times. In contrast, strong vibrational
coupling leads to qualitatively different photon dynamics in the case of
vibrational dressing, which the pure dephasing model cannot capture. Analysis of
the electronic coherence revealed that sufficiently strong coupling to
vibrations (roughly, $S \ge 0.3$) suppresses dynamical superradiance. As the values of $S$ in typical organic molecules are at a maximum around $0.1$, dynamical superradiance should be possible. On the other hand, the maximal values of the order $0.1$ are large enough to see specific effects of the vibrational coupling, such as the asymmetry of the risetime as a function of cavity detuning. 

For experimental studies of the superradiance in the presence of vibrational coupling, the dependence of the risetime on the cavity detuning is interesting. The detuning can typically be easily varied in experiments, and an asymmetry of the risetime as a function of detuning would be a clear signature of a SR process that involves a coherent coupling to vibrational states. In contrast, if a symmetric behavior is observed, it would indicate that the coupling to the vibrational states is not coherent, and they act only as an effective dephasing environment. Another interesting experiment would be to compare the effect of coherent and incoherent pumping. The former is required for the superradiance, and in that case, one would see a more rapid rise in the photon number for smaller inversion, while for incoherent pumping leading to lasing, one would expect the opposite. The time-scales for the parameters we have chosen to use are very small, tens of femtoseconds, but for other parameter regimes, the dynamics could be brought to the regime accessible to state-of-the-art time-resolved techniques. 

The power of the mean-field approaches lies in their independence of
computation complexity from system size, which allows 
for solutions for arbitrarily large \(N\to\infty\), as well
as their broad applicability.  Future work may combine
these and exact 
approaches to study the effects of multiple vibrational modes or a
continuum of modes for the local emitter environments, opening pathways to
explore superradiance and related phenomena across the diverse range of
materials and environments encountered in real physical systems.
\vspace{0.5cm}

\noindent\textbf{Acknowledgments:}
Part of the calculations presented above were performed using computer resources within the Aalto University School of Science “Science-IT” project.

\noindent\textbf{Research funding:}
This work was supported by the Research Council of Finland under project number 339313, and by the Jane and Aatos Erkko Foundation and the Technology Industries of Finland Centennial Foundation as part of the Future Makers funding program.
The work is part of the Research Council of Finland Flagship Programme, Photonics Research and Innovation (PREIN), decision number 346529, Aalto University.

\noindent\textbf{Author contributions:}
All authors have accepted responsibility for the entire content of this
manuscript and consented to its submission to the journal, reviewed all the results and approved
the final version of the manuscript. PT and JK initiated and supervised the research. LF and PFW wrote the code used in the calculations. LF performed the calculations. All authors participated in discussing and interpreting the results. LF prepared the manuscript with
contributions from all co-authors. 

\noindent\textbf{Conflict of interest:}
Authors state no conflicts of interest.

\noindent\textbf{Data availability:}
The datasets generated and analysed during the current study are available in the Zenodo repository at https://doi.org/10.5281/zenodo.17542025.

\appendix
\section{Convergence of mean field approaches}
\label{sec:appendix_convergence}

\begin{figure}
    \centering
    \includegraphics[width=1\linewidth]{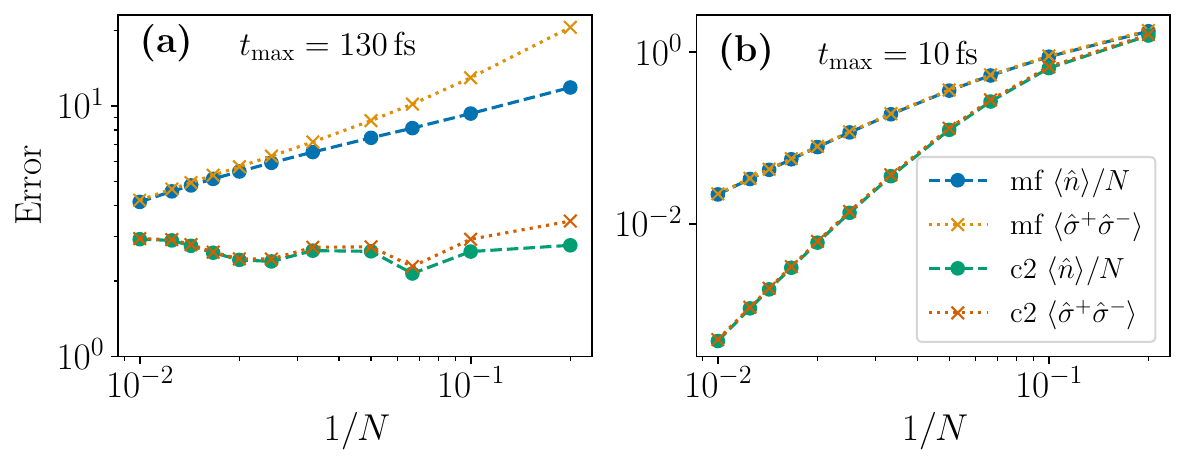}
    \caption{Sum-squared error of the normalized photon number and the two-level system excitation for both mean-field and second-order cumulant solutions as defined in Eq.~\eqref{eq:error} for (a) $t_\text{max}=130\fs$ and (b) $t_\text{max}=10\fs$. For the mean-field solution $\langle \hat n\rangle = |\langle \hat{a}\rangle|^2$ and $\langle\hat\sigma^+\hat\sigma^-\rangle = |\langle \hat\sigma^-\rangle|^2$. The number of two-level systems ranges from $N=5$ to $N=100$ and the superradiance angle is $\theta=\pi/4$. All other parameters are the same as in Fig.~\ref{fig:pibs_c2}.}
    \label{fig:appendix_error}
\end{figure}

In this appendix, we quantify the convergence of mean-field (mf) and second-order cumulant (c2) solutions to the exact solution for the dissipative Tavis-Cummings model from Sec.~\ref{sec:pibs} for superradiance initial conditions ($\theta=\pi/4$). To this end, we define the sum-squared error between the mean-field or cumulant solution and the exact solution for a given quantity $\langle \hat x\rangle$ as
\begin{equation}
    e_\alpha(\hat x) = \sum_{i=1}^{M}(\langle\hat x(t_i)\rangle_\alpha - \langle\hat x(t_i)\rangle_\text{PIBS})^2,\label{eq:error}
\end{equation}
where $\alpha\in\{\text{mf},\text{c2}\}$. The upper bound $M$ of the sum is given by a maximum time $t_\text{max}=t_M$ up to which the error is calculated.
Figure~\ref{fig:appendix_error}(a) shows the so-defined error for the normalized photon number $\langle \hat n\rangle/N$ and for the two-level system population $\langle\hat\sigma^+\hat\sigma^-\rangle$ for different values of $N$ ranging from $5$ to $100$. The maximum time is $t_\text{max}=130\fs$, which covers the entire domain calculated in Fig.~\ref{fig:pibs_c2}. For increasing $N$, the mean-field error monotonically decreases for both calculated quantities. However, this is not the case for the cumulant solution, which does not exhibit a clear convergence behavior. This may at first seem surprising, but the apparent lack of convergence lies in the choice of $t_\text{max}$. As can be seen in Fig.~\ref{fig:pibs_c2}(b) and~\ref{fig:pibs_c2}(d), the cumulant solution agrees well with the exact solution up to some time $t_0$, after which the cumulant solution starts having irregular oscillations. It is in this region $t>t_0$ where a large error can build up in the cumulant solution, because the photon number oscillations can be out of phase with the exact solution. In contrast, the mean-field solution shows regular oscillations in the photon number for all times [cf. Figs.~\ref{fig:pibs_c2}(b), ~\ref{fig:pibs_c2}(d)].
To show the convergence in the initial region $t<t_0$, we set $t_\text{max}=10\fs$ in Fig.~\ref{fig:appendix_error}(b) and calculate the same error quantities as in Fig.~\ref{fig:appendix_error}(a). The error for the photon number and for the two-level system population are equal in that case, and for both the mean-field solution and the cumulant solution a clear convergence behavior is seen for increasing $N$. These results reinforce the main message from Sec.~\ref{sec:pibs}, namely that in the large $N$ limit, the mean-field approximation is reasonable for superradiance initial conditions.

\section{Influence of \texorpdfstring{$\theta$}{θ} on dynamical SR}
\label{sec:appendix}

\begin{figure}
    \centering
    \includegraphics[width=1\linewidth]{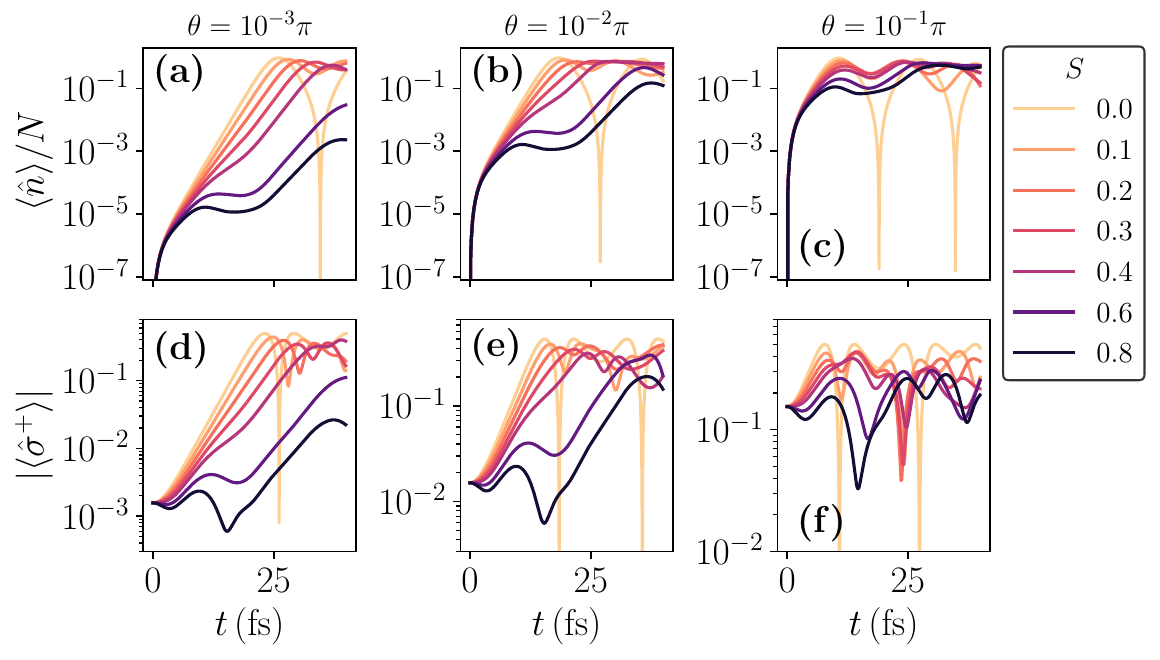}
    \caption{Average number of photons in the cavity mode as a function of time in the HTC model for (a) $\theta=10^{-3}\pi$, (b) $\theta=10^{-2}\pi$, and (c) $\theta=10^{-1}\pi$, for different vibrational coupling strengths. Panels (d)-(f) show the electronic coherence for the same values of $\theta$. All other parameters are the same as in Fig.~\ref{fig:htc_phot}. }
    \label{fig:app_b}
\end{figure}

In this appendix, we investigate the influence of the angle $\theta$ on dynamical SR in the presence of vibrational dressing. All results are calculated with mean-field theory. Figure~\ref{fig:app_b} shows the average number of photons and the electronic coherence as a function of time for three different values of $\theta$ [Figs.~\ref{fig:app_b}(a) and~\ref{fig:app_b}(d) are the same as Figs.~\ref{fig:htc_phot}(a) and~\ref{fig:htc_phot}(b)]. Clearly, an increase in $\theta$ leads to an increase in initial electronic coherence $\vert\langle\hat\sigma^+\rangle\vert = \sin(\theta)/2$, resulting in an increased photon emission rate. However, irrespective of $\theta$, for strong vibrational coupling $S\gtrsim0.4$, the coherence tends to decrease at early times, which indicates that SR is suppressed.

\begin{figure}
    \centering
    \includegraphics[width=1\linewidth]{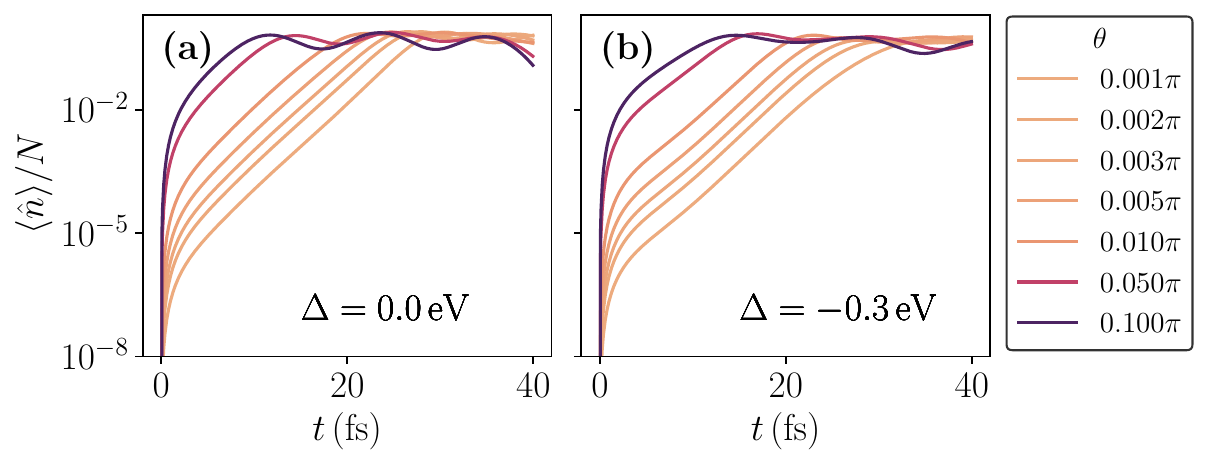}
    \caption{Average number of photons as a function of time in the HTC model for different values of $\theta$ and $S=0.2$, for (a) $\Delta = 0$ and (b) $\Delta = -0.3\eV$. Note the non-uniform spacing of $\theta$. All other parameters are the same as in Fig.~\ref{fig:htc_phot}.}
    \label{fig:htc_theta}
\end{figure}

Figures~\ref{fig:htc_theta}(a) and \ref{fig:htc_theta}(b) show how the angle $\theta$ changes the risetime behavior in the case of vibrational dressing for $\Delta=0$ and $\Delta=-0.3\eV$, respectively. We set $S=0.2$, which ensures that the system still exhibits an exponential rise behavior at early times. We see that for small values of $\theta$, the risetime (defined by fitting $A\ee^{t/\tau}$ to the exponential regions) is approximately constant. However, increasing $\theta$ results in an enhanced initial photon emission, before the exponential regime is reached. If $\theta$ becomes too large, the exponential regime disappears. 


\section{Influence of \texorpdfstring{$g\sqrt{N}$}{g√N} on dynamical SR}
\label{sec:appendix_gSqrtN}
\begin{figure}
    \centering
    \includegraphics[width=1\linewidth]{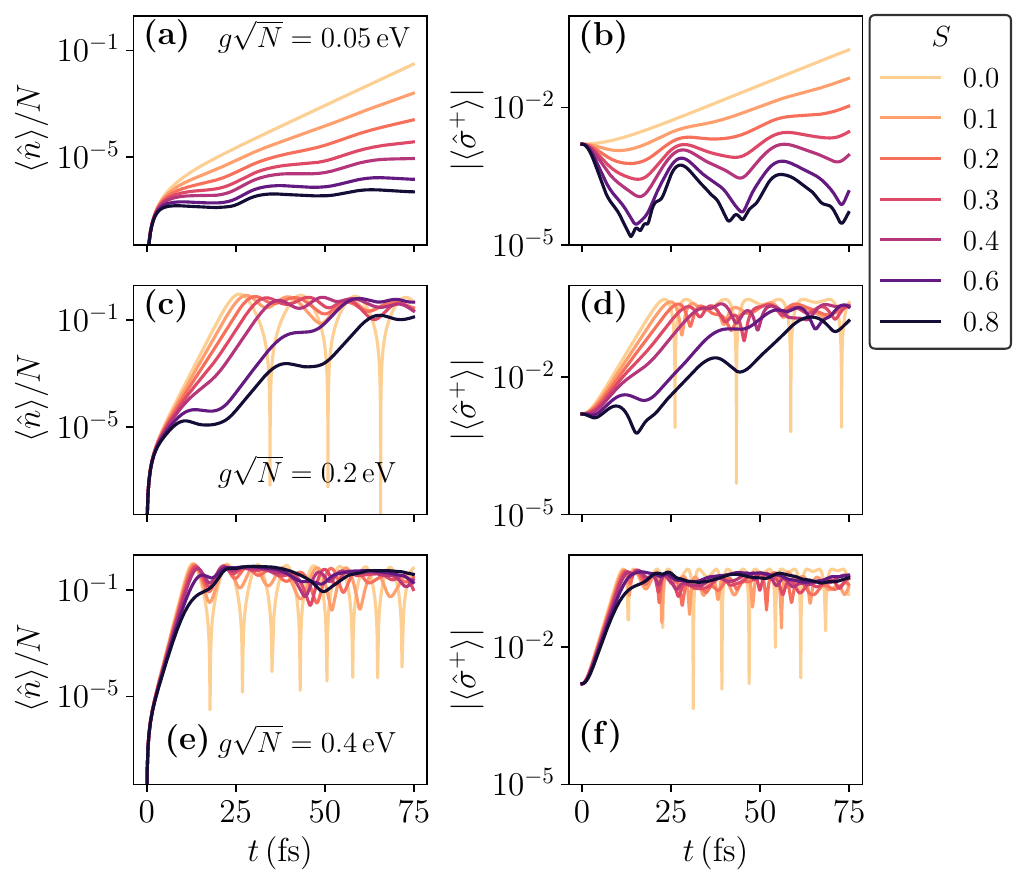}
    \caption{Mean-field solution of the average number of photons $\langle \hat n\rangle/N = |\langle \hat a\rangle|^2/N$ and the electronic coherence $|\langle \hat\sigma^+\rangle|$ in the HTC model for different values of $S$ and $g\sqrt{N}$. (a)-(b): $g\sqrt{N}=0.05\eV$, (c)-(d): $g\sqrt{N}=0.2\eV$ [same as Fig.~\ref{fig:htc_phot}(a)-(b)], (e)-(f): $g\sqrt{N}=0.4\eV$. All other parameters are the same as in Fig.~\ref{fig:htc_phot}(a)-(c).}
    \label{fig:appendix_gSqrtN}
\end{figure}

We study here in more detail the effect of the light-matter coupling strength $g\sqrt{N}$ on the threshold of dynamical SR reported in Sec.~\ref{sec:htc}. To this end, we calculate the same dynamics as in Figs.~\ref{fig:htc_phot}(a) and~\ref{fig:htc_phot}(b) in the main text for three different values of $g\sqrt{N}$, while all other parameters are kept the same. The results are shown in Fig.~\ref{fig:appendix_gSqrtN}, where we note that Figs.~\ref{fig:appendix_gSqrtN}(c) and~\ref{fig:appendix_gSqrtN}(d), corresponding to $g\sqrt{N}=0.2\eV$, are the same as Figs.~\ref{fig:htc_phot}(a) and~\ref{fig:htc_phot}(b). Recall that for $g\sqrt{N}=0.2\eV$, we identified a threshold for dynamical SR for $S\approx0.3$, since beyond that value of $S$, the electronic coherence does not monotonically increase anymore. From Figs.~\ref{fig:appendix_gSqrtN}(b) and~\ref{fig:appendix_gSqrtN}(f), we see that the so-defined SR threshold is at roughly $S\approx0.1$ for $g\sqrt{N}=0.05\eV$, whereas for $g\sqrt{N}=0.4\eV$, even for $S=0.8$ the electronic coherence rises monotonically. We therefore conclude that the SR threshold emerges from a competition between light-matter coupling and vibronic coupling: The stronger the light-matter coupling, the larger the maximum value of $S$ for which dynamical SR is still possible. This makes sense, since $g\sqrt{N}$ corresponds to the Rabi-oscillation frequency of the cavity-emitter system, i.e. larger values of $g\sqrt{N}$ result in a faster initial emission of photons, as seen in Figs.~\ref{fig:appendix_gSqrtN}(a),~\ref{fig:appendix_gSqrtN}(c) and~\ref{fig:appendix_gSqrtN}(e). 

\bibliography{references2.bib}

\end{document}